\documentclass{iaus}
\usepackage{graphicx}

\title[C\,{\sc i} non--LTE spectral line formation in late-type stars] 
{C\,{\sc i} non--LTE spectral line formation in late-type stars}

\author[Fabbian, Asplund, Carlsson and Kiselman]   
{D. Fabbian$^1$,
M. Asplund$^1$, 
M. Carlsson$^2$ \and D. Kiselman$^3$}

\affiliation{
$^1$RSAA, The Australian National University, Mt. Stromlo Observatory,
\break Cotter Rd., Weston ACT 2611, Australia
\break email: damian@mso.anu.edu.au\\[\affilskip]
$^2$Institute of Theoretical Astrophysics, University of Oslo,
\break P.O. Box 1029, Blindern, N-0315 Oslo, Norway\\[\affilskip]
$^3$The Institute for Solar Physics of the Royal Swedish Academy of
Sciences, \break AlbaNova University Centre, 106 91 Stockholm, Sweden
 }
\pubyear{2005}
\volume{228}  
\pagerange{1--2}
\date{?? and in revised form ??}
\jname{From Lithium to Uranium: Elemental Tracers of Early Cosmic Evolution}
\editors{V. Hill, P. Fran\c{c}ois \& F. Primas, eds.}
\begin{document}

\maketitle

\begin{abstract}
We present non--Local Thermodynamic Equilibrium (non--LTE)
calculations for neutral carbon spectral line formation, carried out
for a grid of model atmospheres covering the range of late-type
stars. The results of our detailed calculations suggest that the
carbon non--LTE corrections in these stars are higher than usually
adopted, remaining substantial even at low metallicity. For the most
metal-poor stars in the sample of Akerman et al. (2004), the non--LTE
abundance corrections are of the order of $-0.35...-0.45$ dex (when
neglecting H collisions).  Applying our results to those observations,
the apparent [C/O] upturn seen in their LTE analysis is no longer
present, thus revealing no need to invoke contributions from Pop. III
stars to the carbon nucleosynthesis.
\end{abstract}


The lack of detailed investigations on the non--LTE formation of
spectral lines for a variety of chemical elements is unfortunately a
major obstacle in pushing current stellar abundance studies to the 0.1
dex precision limit or less. It is fundamental to have a clear
understanding of the trends with for example metallicity, for various
elements, as a way to shed light on the chemical evolution of our
Galaxy and of the Universe in general.

We have carried out non--LTE spectral line formation calculations for
C\,{\sc i} using the code {\it Multi} (Carlsson 1986). A grid of MARCS
model atmospheres with stellar parameters in the range
$4500\le$T$_{eff}$$\le7000$, $2.0\le$$\log$ g$\le5.0$ and
$-3\le$[Fe/H]$\le 0$ has been employed. The carbon atomic model used
in this study contains 217 energy levels and 650 radiative transitions
(f-values and photo-ionization cross-sections from the OP database
TOPbase, Cunto et al. 1993) and collisions with electrons and hydrogen
are also included.

The magnitude of the non--LTE effects affecting the formation of
carbon spectral lines in these late-type stars and their trend with
effective temperature, gravity, metallicity and carbon content have
been investigated: spectral lines were found to be generally stronger
in non--LTE across our parameter grid. Thus, with respect to the LTE
approximation, we obtain negative abundance corrections, which become
particularly severe at high temperature (6000-7000 K) and low gravity
(2.0-3.0), where they can reach $\sim -1.0$ dex and more, while for the most
metal-poor halo turn-off stars $\Delta\log\,\epsilon_{\rm C} \simeq-0.35...-0.45$.

The driving non--LTE effects are found to be the source function
dropping below the local Planck function (at solar metallicity), while
increased line opacity is responsible for the non--LTE line
strengthening at low metallicity. The most pronounced corrections are
seen at [Fe/H]$=-1$ dex, where these two effects work in the same
sense. However, corrections are shown to be generally large across the
whole grid.

In particular, to better constrain the evolution of the C/O ratio, we
have focussed our attention on the non--LTE processes affecting the
lines still visible in the metal-poor regime. The influence of
changing the efficiency of H collisions (which probably remain the
major source of uncertainty in our study and in similar ones), has
also been investigated in the present work, showing little sensitivity
of the overall resulting non--LTE abundance corrections for the
metal-poor halo stars of interest, where they become less dramatic but
still remain important (of the order of $\Delta\log\,\epsilon_{\rm C} \simeq
-0.25...-0.35$) when setting efficient hydrogen collisions. Very
recently, statistical equilibrium calculations for carbon across a
grid of stellar parameters similar to the one adopted here, have been
carried out by Takeda \& Honda (2005). These authors find similar
significantly large non--LTE corrections. A detailed comparison with
their results is of interest and, together with full details of our
study, will be shortly available in a subsequent paper (Fabbian et
al. 2005).

\begin{figure}
 \centering \includegraphics[width=10cm]{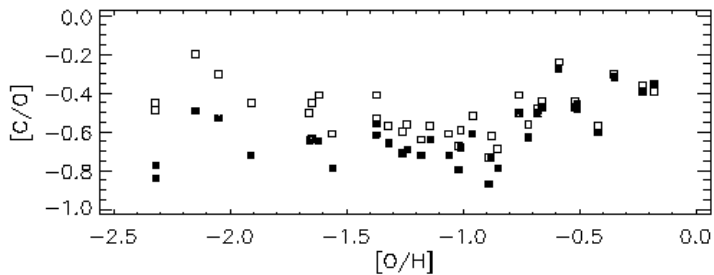}
 \caption{Trend of the [C/O] ratio vs. oxygen abundance, in Milky Way
 halo and disk stars, in logarithmic abundances relative to solar. The
 empty symbols represent literature data (see Akerman et
 al. 2004). Filled symbols represent the abundance values we obtain
 with oxygen non--LTE corrections as adopted by those authors, but
 accounting for our larger non--LTE corrections for carbon. Both
 carbon and oxygen abundance corrections used in this plot are
 calculated neglecting inelastic collisions with hydrogen}
\label{fig:fabbianfig1}
\end{figure}

These findings have important consequences on studies of Galactic
chemical evolution, in particular on the derivation of carbon
abundance at low metallicity. Figure~\ref{fig:fabbianfig1} shows
results of Akerman et al. (2004), where those authors have assumed
that the carbon non--LTE corrections are of the same order as those
calculated for oxygen. An apparent upturn in the [C/O] ratio at low
metallicity is then visible when an LTE analysis is
performed. However, after applying the here calculated non--LTE
corrections, such a trend is no longer present, which suggests it is
probably due to errors intrinsic in the use of the LTE
approximation. Instead, an almost flat ``plateau'' is recovered, with
[C/O]$\sim-0.7$ at [O/H]$=-1$ and remaining approximately constant
down to [O/H]$\sim-2.5$. Thus, there is apparently no need to invoke C
production in Pop. III stars, as advocated by Akerman et al. (2004)
and Spite et al. (2005). More high-quality spectroscopic data for
these metal-poor turn-off stars is highly desirable to better
constrain the trends at low metallicity.


\end{document}